\documentclass[twocolumn,pra,showpacs,nofootinbib,groupedaddress]{revtex4}
\usepackage{amssymb}

\usepackage{dcolumn,graphicx}

\begin{document}

\title{Possibility of an ultra-precise optical clock using the $6\,^1\!S_0
\rightarrow 6\,^3\!P^o_0$ transition in $^{171, 173}$Yb atoms held in an
optical lattice}
\author{Sergey G.~Porsev}
\affiliation {
Department of Physics, University of Nevada, Reno, Nevada 89557 \\
Petersburg Nuclear Physics Institute, Gatchina,
Leningrad district, 188300, Russia}
\author{Andrei Derevianko}
\affiliation {
Department of Physics, University of Nevada, Reno, Nevada 89557}
\author{E.N.~Fortson}
\affiliation {
Department of Physics, University of Washington, Seattle, Washington 98195}

\date{\today}

\begin{abstract}
We report calculations designed to assess the ultimate precision
of an atomic clock based on the 578 nm $6\,^1\!S_0 \rightarrow
6\,^3\!P^o_0$ transition in Yb atoms confined in an optical
lattice trap. We find that this transition has a natural linewidth
less than 10 mHz in the odd Yb isotopes, caused by hyperfine
coupling. The shift in this transition due to the trapping light
acting through the lowest order AC polarizability is found to
become zero at the \textit{magic} trap wavelength of about 752 nm.
The effects of Rayleigh scattering, higher-order polarizabilities,
vector polarizability, and hyperfine induced electronic magnetic
moments can all be held below a mHz (about a part in $10^{18}$),
except in the case of the hyperpolarizability larger shifts due to
nearly resonant terms cannot be ruled out without an accurate
measurement of the magic wavelength.
\end{abstract}


\maketitle

Optical atomic clocks offer new opportunities for creating improved time
standards as well as looking for changes in fundamental constants over time,
measuring gravitational red shifts, and timing pulsars. Compared with
microwave atomic clocks, optical clocks have the intrinsic advantage that
optical transitions have a much higher frequency and potentially much higher
line-Q than microwave transitions. Moreover, optical frequency comb
techniques~\cite{SteSchTam02} now permit different optical frequencies to be
compared with each other at the $10^{-17}$ level or better, and to be linked
to microwave clocks as well. Optical transitions in alkaline earth atoms
offer remarkable possibilities for clocks. In addition to the relatively
sharp $^{1}\!S_{0}\rightarrow \ ^{3}\!P_{1}^{o}$ intercombination line
already in use~\cite{DidUdeBer01}, there occurs the much sharper $%
^{1}\!S_{0}\rightarrow \ ^{3}\!P_{0}^{o}$ line in odd isotopes. This
one-photon transition is forbidden in even isotopes, but in odd isotopes
acquires a weak E1 amplitude induced by the internal hyperfine coupling of
the nuclear spin. Doppler and recoil shifts can be virtually eliminated by
confining very cold atoms in an optical lattice trap. This lattice will be
\textit{Stark-free} if it is produced by laser beams tuned to the \textit{%
magic frequency} where the ground and excited states undergo the same light
shift, leaving the clock transition unshifted and relatively insensitive to
the laser polarization.

Katori~\cite{Kat02} has pointed out these advantages
for $^{87}$Sr, and the $5\,^{1}\!S_{0}\rightarrow 5\,^{3}\!P_{0}^{o}$
transition in this isotope has recently been observed~\cite{CorQueKov03}.
Also, Sr atoms have been held in an optical lattice at the laser wavelength
appropriate for a Stark-free $5\,^{1}\!S_{0}\rightarrow 5\,^{3}\!P_{1}^{o}$
transition, and this transition has been observed free of Doppler and recoil
shifts~\cite{IdoKat03}.
The magic frequency for the Sr clock has
been evaluated recently in Ref.~\cite{KatTakPal03} and it
has been measured in Ref.~\cite{TakKat03}.
In Ref.~\cite{KatTakPal03} an estimate of systematic uncertainties for Sr
also has been carried out.

Ytterbium has two stable odd isotopes, $^{171}$Yb and $^{173}$Yb, which also
appear to be excellent candidates for an atomic standard, using the $%
6\,^{1}\!S_{0}\rightarrow 6\,^{3}\!P_{0}^{o}$ transition at the
convenient wavelength 578 nm. The atoms are readily trapped into a
MOT operating on either the strong $6\,^{1}\!S_{0}\rightarrow
6\,^{1}\!P_{1}^{o}$ line or the $6\,^{1}\!S_{0}\rightarrow
6\,^{3}\!P_{1}^{o}$ intercombination line and in the latter case
have been cooled to $\mu$K temperatures by Sisyphus
cooling~\cite{MarWynRom03}. Also, these isotopes have been
successfully confined in an optical dipole
trap~\cite{HonTakKuw02}. In this paper we present a calculation of
the natural linewidth of the clock transition, which turns out to
be less than 10  mHz, and also a calculation of the Stark-free
wavelength of an optical lattice trap for the clock transition.
This wavelength turns out to be about 752 nm, reachable with
adequate power by a tunable Ti:Sa laser. We also estimate the size
of the polarizability due to higher magnetic dipole and electric
quadrupole optical moments. In addition, we estimate the Rayleigh
and Raman scattering rates in the optical lattice which can limit
the coherence lifetime of the clock transition. Finally, we
compute the small but important hyperfine-induced Zeeman shift in
the excited state and the vector light shift which can cause a
small Stark-shift dependence on the polarization of the trapping
light.

Our results indicate that a light intensity of 10 kW/cm$^2$ would
create a convenient trap depth of 15 $\mu$K at the magic
wavelength, while perturbations to the clock frequency could be
held below the mHz level ($10^{-18}$ relative shift) -- with one
possible exception. Larger shifts due to accidental near
resonances in the hyperpolarizability cannot be ruled out without
an accurate measurement of the magic wavelength.

All the calculations reported in this paper have been carried out using the
relativistic many-body code described in Refs.~\cite
{DzuFlaKoz96b,DzuKozPor98,KozPor99E}. The employed formalism is a
combination of configuration-interaction method in the valence space with
many-body perturbation theory for core-polarization effects. The effective
core-polarization (self-energy) operator is adjusted so that the
experimental energy levels are well reproduced. In addition, the dressing of
the external electromagnetic field (so-called core shielding) is included in
the framework of the random-phase approximation. In the following we refer
to this many-body method as CI+MBPT. For Yb, the CI+MBPT method has an
accuracy of a few per cent for electric dipole matrix elements and
magnetic-dipole hyperfine constants~\cite{PorRakKoz99P,PorRakKoz99J}. Unless
specified otherwise, we use atomic units ($|e|=\hbar=m_e\equiv 1$)
throughout the paper.

\begin{figure}[h]
\begin{center}
\includegraphics*[scale=0.8]{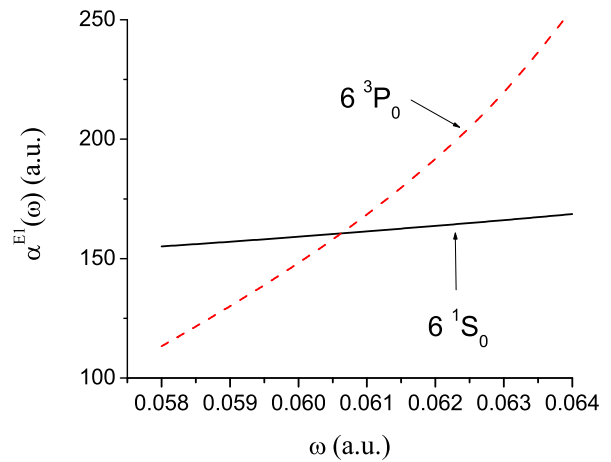}
\end{center}
\caption{ Electric dipole a.c.\ polarizabilities for $6\,^1\!S_0$ (solid
line) and $6\,^3\!P_0$ (dashed line) states of Yb. The polarizabilities are
shown as a function of laser frequency $\protect\omega$. }
\label{Fig:dynPol}
\end{figure}

In the proposed design, the Yb atoms are confined to sites of an optical
lattice (formed by a standing-wave laser field of frequency $\omega $ and
amplitude of electric field $\mathcal{E}_{0}$). To the leading order in
intensity and the fine-structure constant, the laser field shifts the clock
transition frequency $\omega _{0}$ by
\begin{equation}
\delta \omega _{0}(\omega )=-\left\{ \alpha _{6\,^{3}\!P_{0}^{o}}^{E1}\left(
\omega \right) -\alpha _{6\,^{1}\!S_{0}}^{E1}\left( \omega \right) \right\}
\left( \mathcal{E}_{0}/2\right) ^{2}\,,  \label{Eq:omega_pert}
\end{equation}
where $\alpha _{X}^{E1}\left( \omega \right) $ is an a.c.\ electric-dipole
polarizability of level $X$
\begin{equation}
\alpha _{X}^{_{E1}}\left( \omega \right) =2\,\sum_{Y}\frac{E_{Y}-E_{X}}{%
\left( E_{Y}-E_{X}\right) ^{2}-\omega ^{2}}\left| \langle X|D_{z}|Y\rangle
\right| ^{2}\,.  \label{Eq:dynPol}
\end{equation}
We have carried out the calculations of the E1 a.c. polarizability using the
CI+MBPT method. We summed over the intermediate states in Eq.~(\ref
{Eq:dynPol}) using the Dalgarno-Lewis-Sternheimer method~\cite{DalLew55}.
The results of the calculations for both $6\,^{1}\!S_{0}$ and $%
6\,^{3}\!P_{0}^{o}$ states are shown in Fig.~\ref{Fig:dynPol}. The two
dynamic polarizabilities intersect at $\omega ^{\ast }=0.0606$ a.u. At this
``magic'' frequency the lowest-order differential light-shift, Eq.~(\ref
{Eq:omega_pert}), vanishes. It is worth noting that at $\omega ^{\ast }$ the
sum~(\ref{Eq:dynPol}) for the ground state is dominated by the $%
6s6p\,^{1}\!P_{1}^{o}$ state and for the $6\,^{3}\!P_{0}^{o}$ level by
the $6s7s\,^{3}S_{1}$ state.

%Here should come a piece about laser intensity and temperature of the sample.

In general, for a linear laser polarization the second-order light shift of
level $X$ can be represented as a sum over $2^{J}$-pole polarizabilities $%
\alpha _{X}^{\left( J\lambda \right) }$ ($\lambda $ distinguishes between
electric, $\lambda =1$, and magnetic, $\lambda =0$, multipoles)
\begin{equation}
\delta E_{X}=-\frac{\mathcal{E}_{0}^{2}}{4}\sum_{J\lambda }\alpha
_{X}^{\left( J\lambda \right) }\left( \omega \right) \,.
\end{equation}
When the total angular momentum of the level $X$ is equal to zero these a.c.
polarizabilities are expressed as
\begin{eqnarray*}
\alpha _{X}^{\left( J\lambda \right) }\left( \omega \right) &=&\frac{J+1}{J}%
\frac{2J+1}{\left[ \left( 2J+1\right) !!\right] ^{2}}\left( \alpha \omega
\right) ^{2J-2}\times \\
&&\sum_{Y}\left\{ \frac{(E_{Y}-E_{X})\left| \langle Y||Q^{\left( J\lambda
\right) }||X\rangle \right| ^{2}}{\left( E_{Y}-E_{X}\right) ^{2}-\omega ^{2}}%
\right\} \,,
\end{eqnarray*}
with $Q^{\left( J\lambda \right) }$ being relevant multipole operators~\cite
{JohPlaSap95}. Typically, the E1 polarizability~(\ref{Eq:dynPol}) overwhelms
this sum. Compared to the E1 contribution, the higher-order multipole
polarizabilities are suppressed by a factor of $(\alpha \omega )^{2J-2}$ for
EJ and by a factor of $\alpha ^{2}(\alpha \omega )^{2J-2}$ for MJ
multipoles. We verified that at the magic frequency there are no resonant
contributions for the next-order E2 and M1 polarizabilities and we expect $%
\alpha ^{(E2,M1)}\lesssim 10^{-6}\alpha ^{(E1)}$, similar to the case of Sr~
\cite{KatTakPal03}. At the same time we notice that a core-excited state $%
4f^{13}(^{2}\!F_{7/2}^{\circ })5d_{5/2}6s^{2}$ $J=5$ may become
resonant with an excitation from the $6\,^{3}\!P_{0}^{o}$ level. The
relevant $M5$ polarizability is highly suppressed, and we anticipate that
the magic frequency will be only slightly shifted by the presence of this
state.

Higher-order correction to the differential frequency shift, Eq.(\ref
{Eq:omega_pert}), arises due to terms quartic in the field strength $%
\mathcal{E}_{0}$. This fourth-order contribution is expressed in terms of
a.c. hyperpolarizability $\gamma (\omega )$. The expression for $\gamma
(\omega )$~\cite{ManOvsRap86} has a complicated energy denominator structure
exhibiting both single-- and two--photon resonances. While for the ground
state there are no such resonances, for the $6\,^{3}\,P_{0}^{o}$ a
two-photon resonance may occur for $6s8p\,^{1}\!P_{1}^{o}$ and $%
6s8p\,^{3}\!P_{J}^{o}$ intermediate states. Due to theoretical errors in
calculations of the magic frequency we can not reliably predict if the
two-photon resonances would occur. Since the resonance contributions may
dominate $\gamma (\omega )$, we can not provide a reliable estimate of the
fourth-order frequency shift. The estimate may be carried out as soon as the
magic frequency is measured with sufficient resolution. As a possible
indication of the effect on the clock frequency, we notice that for Sr~\cite
{KatTakPal03} the resulting correction to the energy levels was a few mHz at
a trapping laser intensity of 10 kW/cm$^{2}$. This systematic uncertainty
can be controlled by studying the dependence of the level shift on the laser
intensity~\cite{IdoKat03}.

The $6\,^{3}\!P_{0}^{o}$ state decays due to an admixture from $J=1$ states
caused by the hyperfine interaction. In this paper we restrict our attention
to the hyperfine interaction due to the nuclear magnetic moment $\mathbf{\mu
}$. We write this interaction as $H_{\mathrm{hfs}}=\left( \mathbf{\mu}/\mu
_{N}\cdot \mathbf{T}_{e}^{\left( 1\right) }\right) $, where tensor $\mathbf{T%
}_{e}^{\left( 1\right) }$ acts on the electronic coordinates and $\mu _{N}$
is the nuclear magneton. We employ the following nuclear parameters: for $%
^{171}$Yb, the nuclear spin $I=1/2$ and magnetic moment $\mu =0.4919\,\mu
_{N}$, and for $^{173}$Yb, $I=5/2$ and $\mu =-0.6776\,\mu _{N}$. Using
first-order perturbation theory, the HFS-induced transition rate is given by
\begin{equation}
A_{\mathrm{hfs}}\left( 6\,^{3}\!P_{0}^{o}\right) =\frac{4\alpha ^{3}}{27}%
\frac{\left( I+1\right) }{I}\left( \mu /\mu _{N}\right) ^{2}~\omega
_{0}^{3}~\left| S\right| ^{2}\,,  \label{Eq:HFSRate}
\end{equation}
where the sum $S$ is defined as
\begin{equation}
S=\sum_{\gamma ^{\prime }}\frac{\langle 6\,^{1}\!S_{0}||D||\gamma ^{\prime
}\rangle \,\langle \gamma ^{\prime }||\mathcal{T}_{e}^{\left( 1\right)
}||6\,^{3}\!P_{0}^{o}\rangle }{E\left( \gamma ^{\prime }\right) -E\left(
6\,^{3}\!P_{0}^{o}\right) }\,  \label{Eq:S_sum}
\end{equation}
and  $\omega_{0}=E\left( 6\,^{3}\!P_{0}^{o}\right) -E\left(
6^{1}S_{0}\right)$.

To estimate the rate we restricted the summation over intermediate
states to the nearest-energy $6\,^{1}\!P_{1}^{o}$ and
$6\,^{3}\!P_{1}^{o}$ states. Using the CI+MBPT method we computed
HFS couplings, $\langle 6^{3}\!P_{0}^{o}||\mathcal{T}_{e}^{\left(
1\right) }||6\,^{3}\!P_{1}^{o}\rangle =-6685\,\mathrm{MHz}$ and
$\langle 6\ ^{3}\!P_{0}^{o}||\mathcal{T}^{\left( 1\right)
}||6\,^{1}\!P_{1}^{o}\rangle =4019\,\mathrm{MHz}$ and we inferred
the values of dipole matrix elements from lifetime
measurements~\cite{BowBudCom96,TakTakKom03}. The resulting
HFS-induced lifetimes of the $6\,^{3}\!P_{0}^{o}$ level are 20 and
23 seconds for $^{171}$Yb and $^{173}$Yb isotopes respectively.

A coherence of atomic states may be lost due to scattering of laser photons
(Rayleigh and Raman processes~\cite{BerLifPit82}). These are second order
processes. In particular, the Rayleigh (heating) rate for both $%
6\,^{3}\!P_{0}^{o}$ and the ground states may be expressed in terms of a.c.
polarizability
\[
\gamma _{h}=\alpha ^{4}\,\frac{8\pi }{3}\,(\omega ^{\ast })^{3}\,\left[
\alpha ^{E1}\left( \omega ^{\ast }\right) \right] ^{2}I_{L}\,,
\]
where $I_{L}$ is the intensity of laser. At the magic frequency $\omega
^{\ast }$ the values of a.c. polarizability for both states are equal to 160
a.u. (see Fig.~\ref{Fig:dynPol}). For a laser intensity of 10 kW/cm$^{2}$,
the resulting rate is in the order of $10^{-3}$ sec$^{-1}$. As to the
Raman rates, there are no Raman transitions originating from the ground
state. The final states for transitions from the $6\,^{3}\!P_{0}^{o}$ are
the $J=1,2$ sublevels of the same $6\,^{3}\!P_{J}^{o}$ fine-structure
multiplet. We estimate this rate by approximating the relevant second-order
sum with the dominant contribution from the $6s7s\,^{3}\!S_{1}$ intermediate
state. The resulting Raman scattering rate is also in the order of $10^{-3}$
sec$^{-1}$ for 10 kW/cm$^{2}$ laser.

The total magnetic moment of the Yb atom is composed of the nuclear and
electronic magnetic moments. Disregarding shielding of externally applied
magnetic fields by atomic electrons, the $g$-factor due to the nuclear
moment is given by $\delta g_{\mathrm{nuc}}=-(1/m_{p})(\mu /\mu _{N})/I$,
where $m_{p}$ is the proton mass. The numerical values of $\delta g_{\mathrm{%
nuc}}$ are $-5.4\times 10^{-4}$ for $^{171}$Yb and $1.48\times 10^{-4}$ for $%
^{173}$Yb. The electronic magnetic moment of the
$6\,^{3}\!P_{0}^{o}$ state arises due to mixing of levels caused
by the hyperfine interaction, i.e., the same mechanism that causes
the $6\,^{3}\!P_{0}^{o}$ state to decay radiatively. This
correction may be expressed as
\[
\delta g_{\mathrm{hfs}}\approx  \frac{\sqrt{8}}{3}\;\frac{1}{I}\frac{\mu }{%
\mu _{N}}\frac{\langle 6\,^{3}\!P_{1}^{o}||\mathcal{T}_{e}^{\left( 1\right)
}||6\,^{3}\!P_{0}^{o}\rangle }{E\left( 6\,^{3}\!P_{1}^{o}\right) -E\left(
6\,^{3}\!P_{0}^{o}\right) }\,.
\]
The computed values of the $\delta g_{\mathrm{hfs}}$ correction are $%
2.9~\times 10^{-4}$ for $^{171}$Yb and $-8.1\times 10^{-5}$ for $^{173}$Yb,
which imply that mHz shifts would be produced by $\mu $G magnetic fields.
Fields can readily be calibrated and stabilized to this level using magnetic
shielding.

The hyperfine interaction also induces residual vector (axial) $\alpha
_{\gamma F}^{A}\left( \omega \right) $ and tensor $\alpha _{\gamma
F}^{T}\left( \omega \right) $ a.c. polarizabilities. For $J=0$ levels there
is no tensor contribution for the $^{171}$Yb isotope ($I=1/2$) and it can be
shown that for the $^{173}$Yb ($I=5/2$) it vanishes when the HFS interaction
is restricted to the dominant magnetic-dipole term. For a non-zero degree of
circular polarization $\mathcal{A}$, the relevant correction to the light
shift of level $\gamma F$ is
\begin{equation}
\delta E_{\gamma F}^{(A)}=-\frac{M}{2F}\mathcal{A}~\alpha _{\gamma
F}^{A}\left( \omega \right) \left( \frac{1}{2}\mathcal{E}_{0}\right) ^{2}\,,
\end{equation}
where for $J=0$, $F=I$ and $M$ is the magnetic quantum number. Using
third-order perturbation theory and a formalism of quasi-energy states~\cite
{ManOvsRap86} we arrived at an expression for $\alpha _{\gamma F}^{A}\left(
\omega \right) $ which contains two dipole and one hyperfine operator in
various orderings and double summations over intermediate states. Analyzing
these expressions, we find that the vector polarizability of the $%
6^{3}\!P_{0}^{o}$ state is much larger than that for the ground
state, as in the case of Sr~ \cite{KatTakPal03}. For Sr,
\citet{KatTakPal03} estimated the vector polarizability by adding
HFS correction to the energy levels of intermediate states in
Eq.~(\ref{Eq:dynPol}). Our analysis is more complete and we find
that the dominant effect is not due to corrections to the energy
levels, but it is rather due to perturbation of the $6^{3}\!P^o_0$
state by the HFS operator. The resulting values of $\alpha
_{6^{3}\!P^o_0}^{A}\left( \omega ^{\ast }\right) $ are $-0.10$
a.u. for $^{171}$Yb and $0.075$ a.u. for $^{173}$Yb. Using these
values in the above equation, we find that holding $\mathcal{A}$
to $<10^{-6}$ with fields of 10 kW/cm$^{2}$ would keep shifts in
the clock frequency below the mHz level. This requirement on
optical polarization is not an extreme one, and in the special
case of a 1D optical lattice could be relaxed significantly by
orienting the quantization axis (defined by the external magnetic
field) perpendicular to the trap axis.

In conclusion, we have analyzed the possibility of creating a highly
precise optical clock operating on the $6\,^1\!S_0 \rightarrow 6\,^3\!P_0^{o}$
transition in odd isotopes of atomic Yb.  According to our calculations,
the natural linewidth is about 10 mHz, and the magic wavelength for
producing zero Stark shift of this transition in an optical lattice trap is
about 752 nm.  We have examined possible sources of shifts and broadening
due to both the optical trapping fields and any magnetic fields, and find
they should not perturb the clock above the $10^{-18}$ level, except for
possible larger near-resonant terms in the hyperpolarizability.  An
accurate measurement of the magic wavelength will be needed to settle this
last question.

This work was partially supported by the National Science Foundation, grants
PHY 0099535 and PHY 0099419. The work of S.G.P. was partially
supported by the Russian Foundation for Basic Research under grant No
02-02-16837-a.

%\bibliographystyle{plain}
%\bibliography{clocks}

\end{document}